\documentclass[%
 reprint,superscriptaddress,groupedaddress
 amsmath,amssymb,
 aps]{revtex4-2}

\usepackage{graphicx}
\usepackage{dcolumn}
\usepackage{bm}
\usepackage{siunitx}
\usepackage{braket}
\usepackage{amsmath,amssymb,amsfonts}
\usepackage{hyperref}
\hypersetup{colorlinks=true,allcolors=blue}
\usepackage[table]{xcolor} 
\usepackage{booktabs} 
\definecolor{lightgray}{gray}{0.9}
\usepackage{todonotes}

\newcommand{\UIBK}{Institute f{\"u}r Experimentalphysik, Universit{\"a}t Innsbruck, Innsbruck, Austria}
\newcommand{\WWU}{Institut f{\"u}r Festk{\"o}rpertheorie, Universit{\"a}t M{\"u}nster, 48149 M{\"u}nster, Germany}
\newcommand{\TUdo}{Condensed Matter Theory, Department of Physics, TU Dortmund, 44221 Dortmund, Germany}
\newcommand{\Bayreuth}{Theoretische Physik III, Universit{\"a}t Bayreuth, 95440 Bayreuth, Germany}

\begin{document}

\preprint{arxiv}

\title{Theory of time-bin entangled photons from quantum emitters}

\author{Thomas K. Bracht}%
    \affiliation{\TUdo}
    \email{thomas.bracht@tu-dortmund.de}
    \affiliation{\WWU}
\author{Florian Kappe}
    \affiliation{\UIBK}
\author{Moritz Cygorek}
    \affiliation{\TUdo}
\author{Tim Seidelmann}%
    \affiliation{\Bayreuth}
\author{Yusuf Karli}
    \affiliation{\UIBK}
\author{Vikas Remesh}%
    \affiliation{\UIBK}
\author{Gregor Weihs}%
    \affiliation{\UIBK}
\author{Vollrath Martin Axt}%
    \affiliation{\Bayreuth}
\author{Doris E. Reiter}%
    \affiliation{\TUdo}

\date{\today}

\begin{abstract}
Entangled photon pairs form the foundation for many applications in the realm of quantum communication. For fiber-optic transfer of entangled photon pairs, time-bin encoding can potentially offer an improved stability compared to polarization encoded qubits. Here, we lay the theoretical foundations to describe the measurement of time-bin entangled photons. We derive multi-time correlation functions of the time-bin encoded photon pairs, corresponding to quantum state tomographic measurements. Our theory can be the starting point to extend the simulations to include all kinds of loss or decoherence effects that apply in a specific quantum system for realistic simulation for time-bin entanglement from quantum emitters. 
\end{abstract}

\maketitle

\section{Introduction}
\noindent{}The effect of entanglement is a fundamental quantum mechanical property, which lacks a direct classical analog. In an entangled system, a measurement performed on one sub-system directly affects the measurement on the other sub-systems. Entanglement can be found in a multitude of different systems, but for quantum information technology including quantum communication, the usage of photons is the logical choice for ``flying qubits'', as they can be sent through free space or coupled to optical fibers, such that existing infrastructures can be used \cite{vajner2022quantum}. Depending on which degree of freedom of the photons is measured, different types of entanglement can be distinguished. Probably best-known is the polarization entanglement of photons \cite{orieux2017semiconductor,seidelmann2022two}, where the polarization of each of the photons is measured. Similarly to how polarization measurements categorize photons into, for instance, horizontally and vertically polarized photons, partitioning based on other degrees of freedom is also possible, for example the photon number \cite{wein2022photonumber}. Here, we focus on the entanglement based on the emission time of the photons, which is called time-bin entanglement \cite{tittel2001photonic}. \\
Optical fibers are usually not polarization-maintaining. Thus, sending polarization entangled photons becomes problematic when bridging longer distances or the fibers are subject to environmental impacts such as mechanical vibrations or different temperatures. In contrast, time-bin entanglement does not suffer from such effects in fibers, as the time separation is usually much shorter than the time scale on which the environment influences the fiber. Hence, sending time-bin encoded photon pairs through fibers is more robust.
Several works focus on time-bin entanglement from photon pair sources \cite{takesue2005generation,xing14experimental,thiel2023timebin} and recently, specifically from semiconductor quantum dots \cite{jayakumar2014timebin,pril2018hyper,gines2021cavity} with first theoretical approaches to describe it \cite{simon2005creating,pathak2011coherent,brecht2015photon,tiurev2021fidelity}. 

In theory, describing time-bin entanglement and the corresponding measurements is non-trivial, as a quantization of the time-axis as well as the tomographic measurements is less obvious. Here, we give a step-by-step derivation of the equations for the multi-time correlation function to model the quantum state tomographic measurements \cite{altepeter2005photonic,james2001measurement}. Our results are the foundation to simulate time-bin entanglement from quantum emitters, including the microscopic description of the interaction with the environment \cite{cygorek2022simulation}. 

\section{Sectioning into Time-Bins}
\begin{figure}[t]
    \centering
    \includegraphics{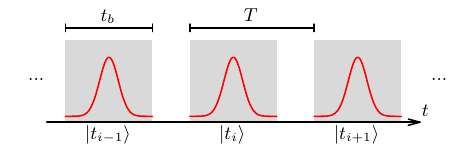}
    \caption{
    \label{fig:timebins}
    Partitioning of the time axis into time-bins that start at a time $t_i$. The photons are contained in the time-bins of width $t_b$, that are separated by a time $T$.
    }
\end{figure}
\noindent{}In time-bin encoding of photons, time not only defines the order of different processes and actions taken on the quantum emitter, such as the atom, molecule or quantum dot \cite{senellart2017high}. Here, it also acts as a degree of freedom, like the polarization in case of polarization entanglement, forming an orthonormal basis in the state-space. The time axis is partitioned into several time-bins of width $t_b$, each separated by a time interval $T$, as shown in Fig.~\ref{fig:timebins}. It is crucial that the time-bins are well separated (i.e., $T\ge t_b$), as this allows them to be understood as an orthogonal basis of states. Photons detected during a time-bin $\ket{t_i}$ can then be distinguished from photons arriving in earlier or later time-bins. 
In contrast to other degrees of freedom, the basis of time-bins offers an infinite number of basis states $\ket{t_i}$ that can also be used for more complex scenarios like the generation of multipartite entanglement \cite{wein2022photonumber,tiurev2021fidelity,bauch2024time} or time-bin based boson sampling \cite{lubasch2018tensor,dhand2018proposal}.\\
For the entanglement between two photons, it is sufficient to focus also on two time-bins, constituting the logical states $\ket{0}$ and $\ket{1}$. In the context of time-bin encoding, these are called the early ($\ket{E}$) and late ($\ket{L}$) time-bins. When looking at Fig.~\ref{fig:timebins}, the early time-bin $\ket{E}$ can be set to an arbitrary $\ket{t_i}$, spanning the time $t\in\left[t_i,t_i+t_b\right]$ and the late time-bin to $\ket{L} = \ket{t_{i+1}}=\ket{t_i+T}$. For simplicity, we define $\ket{E} = \ket{t_0=0}$ and $T=t_b$, leaving no separation between these two time-bins.\\
For a source that emits two photons in these time-bins, a perfectly entangled state will read
\begin{equation}
    \ket{\psi} = \frac{1}{\sqrt{2}}\left(\ket{E}_S\ket{E}_I \pm \ket{L}_S\ket{L}_I\right), \label{eq:entangled_state}
\end{equation}
where the indices $S,I$ denote the signal and idler photon that are distinguishable and therefore separable, for example by their frequency, polarization or spatial mode.\\ 
Experimentally, the state of time-bin encoded photons can be retrieved using quantum state tomography, which relies on coincidence measurements with a delay line in an unbalanced interferometer as shown in Fig.~\ref{fig:onephoton_measurement_setup}(a), interfering the early and late photon states \cite{takesue2005generation,jayakumar2014timebin}. The longer path in the analyzing interferometer introduces a delay matching the time difference between the creation of early and late photons, which is the time-bin separation. Consequently, an early photon traversing the long path in the interferometer becomes indistinguishable from a photon in the late time-bin taking the short path. Coincidences are then measured between the exciton and biexciton photons, usually also with respect to a trigger pulse to mark the beginning of the first time-bin.

\section{Single time-bin encoded photons}\label{sec:classification}
\noindent{}The theoretical description of time-bin entanglement is a delicate subject, because it is not a priori clear what the correct quantization basis is and whether the measured quantity describes a density matrix. 

Before considering the two-photon entanglement, we look at the results of the measurement process in a simplified version by focusing on a single photon emitted in a superposition state given by $\ket{\psi} = \frac{1}{\sqrt{2}}\left(\ket{E} + \ket{L}\right)$. 

\begin{figure*}[t]
    \centering
    \includegraphics[width=\textwidth]{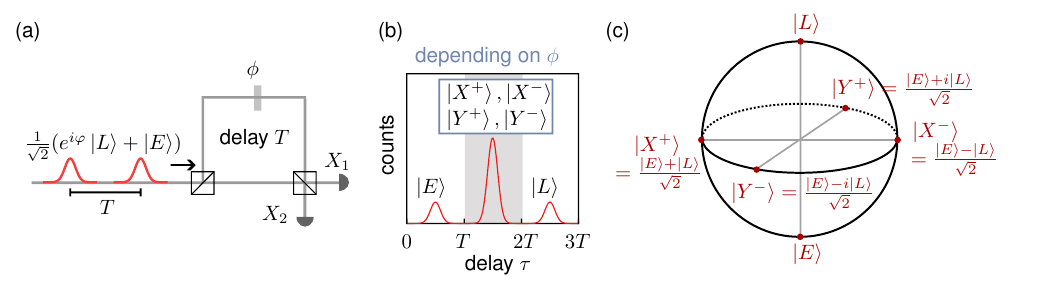}
    \caption{
    (a) Measurement setup for the detection of a time-bin encoded photon.
    The signal is sent through an unbalanced Mach-Zehnder interferometer composed of two beam splitters, where the long arm induces a delay $T$ corresponding to the time-bin separation and an additional phase $\phi$ can be set. The signal is then detected at the detectors $X_1,X_2$. (b) Schematic illustration of a measurement of coincidence counts between trigger pulse and signal. (c) Time-bin states on the Bloch sphere.
    }
    \label{fig:onephoton_measurement_setup}
\end{figure*}

The single photons that are in the superposition of early and late time-bin enter the measurement set-up schematically shown in Fig.~\ref{fig:onephoton_measurement_setup}(a). The photons are then routed through an unbalanced Mach-Zehnder interferometer. The first beam splitter (BS) sends half of the signal to the long arm of the interferometer, which induces a time delay $T$ that matches the time-bin separation. Additionally a phase plate with phase $\phi$ is inserted in the long path. The signals of both arms are recombined using a second beam splitter and routed to two detectors $X_1,X_2$. 

\subsection{Quantum state tomography}
\label{sec:1ph_tomo}
\noindent{}Quantum state tomography can be used to find the density matrix of a quantum system. In experiments, this requires separate measurements done in different bases. Afterwards, the recorded counts of these measurements can be used to reconstruct the density matrix \cite{james2001measurement, altepeter2005photonic}. We simulate this experiment, here for the simplest case of a single photon. In other words, we model the tomography of a single qubit consisting of the states $\ket{E},\ket{L}$. \\
The corresponding density matrix is expanded into the basis of the Pauli matrices $\sigma_j$:
\begin{equation}
    \rho = \begin{pmatrix}
        \rho_{EE} & \rho_{EL}\\
        \rho_{LE} & \rho_{LL}
    \end{pmatrix} = 
    \frac{1}{2}\sum_{j=0}^{3} S_j \sigma_j.\label{eq:QST_DM}
\end{equation}
The three real Stokes parameters $S_i= \mathrm{Tr}(\sigma_i \rho)$ can be determined by measurements in different bases. In total, measurements in three different bases need to be performed in order to obtain the location of the state on the Bloch sphere as shown in Fig~\ref{fig:onephoton_measurement_setup}(c). Besides the distinct time-bin states  $\ket{E},\ket{L}$, we mark the equal superposition states as $\ket{\Phi}=\tfrac{1}{\sqrt{2}}\left( \ket{E}+ e^{i\phi}\ket{L} \right)$, with the distinct states $\ket{X^{\pm}}=\tfrac{1}{\sqrt{2}}\left( \ket{E}\pm\ket{L} \right)$ and $\ket{Y^{\pm}}=\tfrac{1}{\sqrt{2}}\left( \ket{E}\pm i \ket{L} \right)$. Note that in principle any basis consisting of three states $\ket{\tilde{\psi}}$ can be used for the tomography, as long as the matrices $\ket{\tilde{\psi}}\bra{\tilde{\psi}}$ and the identity are linearly independent \cite{altepeter2005photonic}. An unequal superposition state can be achieved by replacing the first beam-splitter with one having a variable splitting ratio. 
\\
In the one-qubit case we can rewrite the equations using $\braket{\ket{E}\bra{E}} + \braket{\ket{L}\bra{L}} = 1$ to
\begin{align}
\begin{split}
    S_0 &= \mathrm{Tr}\left[(\ket{E}\bra{E}+\ket{L}\bra{L})\rho\right] = 1 \, \, (\mathrm{normalization})\\
    S_1 &= \mathrm{Tr}\left[(\ket{X^+}\bra{X^+}-\ket{X^-}\bra{X^-})\rho\right]\\
    &= 2\braket{\ket{X^+}\bra{X^+}} - 1\\
    S_2 &= \mathrm{Tr}\left[(\ket{Y^+}\bra{Y^+}-\ket{Y^-}\bra{Y^-})\rho\right]\\
    &= 2\braket{\ket{Y^+}\bra{Y^+}}-1\\
    S_3 &= \mathrm{Tr}\left[(\ket{E}\bra{E}-\ket{L}\bra{L})\rho\right] = 2\braket{\ket{E}\bra{E}}-1\label{eq:QST_stokes}
\end{split}
\end{align}
In experiments,  the expectation values in Eq.~\eqref{eq:QST_stokes} correspond to detector counts that are collected for different settings of the phase $\phi$ using the phase plate in the measurement interferometer shown in Fig.~\ref{fig:onephoton_measurement_setup}(a). 

In general, the time-resolved detection events of a state $\ket{\psi} \sim \ket{E} + e^{i\varphi}\ket{L}$ reveal three peaks as schematically shown in Fig.~\ref{fig:onephoton_measurement_setup}(b). The first peak corresponds to an early photon taking the short path of the interferometer, while the last is a late photon taking the long path. These results correspond to measurements in the time-basis, i.e., to $\braket{\ket{E}\bra{E}}$ and $\braket{\ket{L}\bra{L}}$, respectively. Early photons taking the long path or late photons taking the short path show up in the interference peak in the middle. This corresponds to a projection onto the basis states $\ket{X^{\pm}},\ket{Y^{\pm}}$, depending on the phase $\phi$ that is set in the interferometer. For example, for $\phi=0$ the measurement is proportional to $\braket{\ket{X^+}\bra{X^+}}$.\\

For a single photon, a coincidence measurement between the two detectors is not necessary, but coincidences are detected with respect to a trigger pulse, marking the start of the time-bin. The output between the two detectors just differs by a constant phase and thus, the counts of the second detector do not hold new information. However, when restricting the analysis to only a single detector, half the counts are missing, which one needs to account for during the reconstruction of the density matrix \cite{takesue2005generation}.\\

\subsection{Simulation of tomographic measurements and reconstruction}
\label{sec:1ph_sim}
\noindent{}To simulate the tomography, the expectation values in Eq.~\eqref{eq:QST_stokes} need to be calculated. Starting with the detector counts, in theory these are described by the first order correlation function $G^{(1)}(t) = \braket{a^\dagger(t)a(t)}$. Here, $a^{\dagger}$ and $a^{}$ are the photon creation and annihilation of a single output mode of the beam splitter. In general, photons in a beam splitter can be described by two input and two output modes. Given the above arguments, it is sufficient to focus on one mode.\\

To account for the time delay introduced by the unbalanced interferometer and the phase $\phi$ introduced in the long path, we describe the mode as \cite{pathak2011coherent}
\begin{equation}
    a(t)\rightarrow a(t) + e^{i\phi}a(t-T) \,.
\end{equation}
With this separation the first order correlation function reads
\begin{widetext}    \begin{subequations} \begin{align} \label{eq:1pulseg1}
    G^{(1)}(t,\phi) &= 
        \braket{a^{\dagger}(t)a(t)} 
        + \braket{a^{\dagger}(t-T)a(t)}e^{-i\phi} + \braket{a^{\dagger}(t)a(t-T)}e^{i\phi}
        + \braket{a^{\dagger}(t-T)a(t-T)} \\
    &=\braket{a^{\dagger}(t)a(t)} 
        + 2\cos(\phi-\varphi)|\braket{a^{\dagger}(t)a(t-T)}|
        + \braket{a^{\dagger}(t-T)a(t-T)} .\label{eq:g1_visibility}
\end{align} \end{subequations}
\end{widetext}
Here, $\varphi$ is the relative phase between early and late state in the input photon state and appears in the interference term.\\
To experimentally measure this quantity, it is correlated with a trigger pulse marking the first time-bin to account for finite measurement times \cite{pathak2011coherent}
\begin{equation}
     G^{(2)}(\tau,\phi) = \int_0^\infty\! dt\, |\Omega_0(t)|G^{(1)}(t+\tau,\phi)    \,.
\end{equation}
By this, a relative time-axis is defined in experiments. It can be understood as a coincidence measurement where $\Omega_0(t)$ is a classical trigger pulse. When calculated, a signal consisting of the three distinct peaks spaced by the time-bin duration $T$ as indicated in Fig.~\ref{fig:onephoton_measurement_setup}(b) will occur, as discussed above. For the tomography, the counts are retrieved by integrating over these peaks in the signal. 

In the theoretical model, we can assume infinitely fast detectors and can therefore work directly with the $G^{(1)}$-function. In this case, the tomographic measurements translate to an integral of $G^{(1)}(t)$ over the respective time-bins.\\
For the pulsed excitation we consider the case that the system is only excited during the two time-bins. We assume that after each excitation, the system relaxes during the period of one time-bin (i.e., emission does not leak into a next time-bin). That gives a time window $[0:2T]$ where emission from the quantum emitter is present. Inspecting Eq.~\eqref{eq:1pulseg1}, the three terms are non-vanishing in different time windows. The first term $~ \braket{a^{\dagger}(t)a(t)} $ is non-vanishing for $[0:2T]$ and due to the shift the last term $~ \braket{a^{\dagger}(t-T)a(t-T)} $ is non-vanishing  only for $[T:3T]$. The middle term $~ \braket{a^{\dagger}(t)a(t-T)} $ is only non-vanishing in the window $[T:2T]$. By slicing the integral into three time windows, the counts for four of our distinct basis states then read

\begin{widetext}
    \begin{align}
     \int_0^{3T} G^{(1)}(t,\phi) dt &= 
        \underbrace{\int_0^{T} G^{(1)}(t,\phi) dt}_{P_{\ket{E}\bra{E}}}
        +\underbrace{\int_T^{2T} G^{(1)}(t,\phi) dt}_{
            \substack{P_{\ket{\Phi}\bra{\Phi}}}}
        +\underbrace{\int_T^{3T} G^{(1)}(t,\phi) dt}_{P_{\ket{L}\bra{L}}} \qquad \text{with}
    \end{align}
    \begin{subequations}
    \begin{align}
    P_{\ket{E}\bra{E}} &=\int_0^T \!dt\,\braket{a^{\dagger}(t)a(t)}.\\
    P_{\ket{X^+}\bra{X^+}} &= \int_{T}^{2T}\! dt\, \braket{a^{\dagger}(t)a(t)} + \braket{a^{\dagger}(t-T)a(t)} + \braket{a^{\dagger}(t)a(t-T)} + \braket{a^{\dagger}(t-T)a(t-T)}
    \propto \braket{\ket{X^+}\bra{X^+}},\label{eq:xplus}\\
    P_{\ket{Y^+}\bra{Y^+}} &= \int_{T}^{2T}\! dt\, \braket{a^{\dagger}(t)a(t)} + \braket{a^{\dagger}(t-T)a(t-T)} -i \braket{a^{\dagger}(t-T)a(t)} +i \braket{a^{\dagger}(t)a(t-T)} \propto \braket{\ket{Y^+}\bra{Y^+}}.\label{eq:yplus}\\
    P_{\ket{L}\bra{L}} &=\int_{2T}^{3T} \!dt\,\braket{a^{\dagger}(t-T)a(t-T)}.
\end{align} \label{eq:onephoton_P}\end{subequations}
\end{widetext}
Due to the action of the operators in the different times, we find that only in the middle time window all expectation values contribute. 
Comparing Eq.~\eqref{eq:onephoton_P}(b-c) 
with
 \begin{subequations}\begin{align}
\begin{split}
    \braket{\ket{X^+}\bra{X^+}} &= \frac{1}{2}(\braket{\ket{E}\bra{E}} + \braket{\ket{L}\bra{L}}\\
    &+ \braket{\ket{E}\bra{L}} + \braket{\ket{L}\bra{E}}) 
\end{split}\\
\begin{split}
        \braket{\ket{Y^+}\bra{Y^+}} &= \frac{1}{2}(\braket{\ket{E}\bra{E}} + \braket{\ket{L}\bra{L}}\\
    &-i \braket{\ket{E}\bra{L}} +i \braket{\ket{L}\bra{E}}),
\end{split}
\end{align} \end{subequations}
and using Equations~\eqref{eq:QST_DM}-\eqref{eq:QST_stokes}, we can calculate the elements of the (at this point not normalized) density matrix using
 \begin{subequations}\begin{align}
    \tilde{\rho}_{EE} &= \braket{\ket{E}\bra{E}} = \int_0^T\!dt\, \braket{a^{\dagger}(t)a(t)} \\
    \tilde{\rho}_{EL} &= \braket{\ket{L}\bra{E}} = \int_T^{2T}\!dt\, \braket{a^{\dagger}(t)a(t-T)} \\
    \tilde{\rho}_{LE} &= \braket{\ket{E}\bra{L}} = \int_T^{2T}\!dt\, \braket{a^{\dagger}(t-T)a(t)}\\
    \tilde{\rho}_{LL} &= \braket{\ket{L}\bra{L}} = \int_{2T}^{3T}\!dt\, \braket{a^{\dagger}(t-T)a(t-T)}
\end{align} \end{subequations}
Introducing the early and late operators $a_E(t) = a(t), a_L(t) = a(t+T)$, we can summarize the calculation of the normalized density matrix as
\begin{subequations}\begin{align}
    &\rho_{j,k} = \frac{\overline{G}^{(1)}_{j,k}}{\text{Tr}\left\{\overline{G}^{(1)}\right\}},\\
    &\overline{G}^{(1)}_{j,k} = \int_0^T\! dt\, \braket{a_k^{\dagger}(t)a_j(t)}, \quad j,k\in\{E,L\}.
\end{align} \end{subequations}
In the experiment, the phase plate is rotated and the visibility, i.e., the ratio between sum and difference between minimal and maximal counts, is measured. To understand the phase-sensitivity of the measurement, we take a second look at Eq.~\eqref{eq:g1_visibility} to find for the counts in the middle peak  
\begin{equation}
    \int_{T}^{2T}\!\!dt\,G^{(1)}(t,\phi) \propto \rho_{EE} +\rho_{LL} + 2\cos(\phi-\varphi)|\rho_{EL}|.\label{eq:g1_middle_peak}
\end{equation}
Turning the phase plate $\phi$ yields a cosine, with the maximal (minimal) counts being proportional to $\rho_{EE}+\rho_{LL}\pm 2|\rho_{EL}|$, resulting in a visibility of $V=2|\rho_{EL}|$. For an ideal superposition state, $|\rho_{EL}|=0.5$ and the visibility is unity, while any deviations from the ideal state result in a loss of visibility. However, Eq.~\eqref{eq:g1_middle_peak} also reveals the importance of using a phase-stable interferometer as well as phase-locked pulses: a change in the phases for subsequent measurements in addition to the impact of the phase plate will also result in a reduced visibility.
\begin{figure*}[t]
    \centering
    \includegraphics[width=0.95\textwidth]{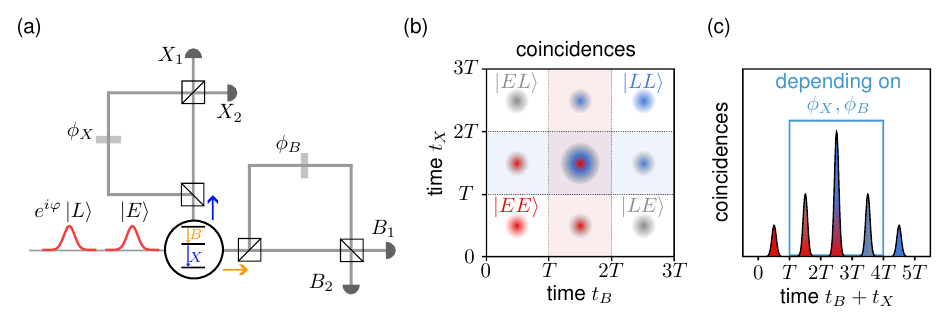}
    \caption{
    (a) Measurement setup for the detection of time-bin entangled photon pairs. The pair source (here a quantum dot) is excited to emit the entangled state $\ket{\psi} = \frac{1}{\sqrt{2}}\left(\ket{EE} + \ket{LL}\right)$. Before coincidence detection, the emitted photons are routed through unbalanced Mach-Zehnder interferometers that add a delay equal to the time-bin separation if a photon traverses their long arm, while a phase plate can be used to imprint an additional phase onto the photon. (b) Schematic picture of time-resolved coincidences measured between biexciton and exciton channel. The corners of the $3\times3$ histogram correspond to the respective combinations of $E$ and $L$, while the remaining peaks are interferences between the neighboring states. (c) Coincidences depending on the arrival time $t=t_B+t_X$. This corresponds to a diagonal projection of the diagram shown in (b) 
    }
    \label{fig:full_measurement_setup}
\end{figure*}
\section{Photon pair states}
\noindent{}Now we apply the same procedure we have just discussed for a single photon to the two-photon states, which can be in the time-bin entangled state as given in Eq.~\eqref{eq:entangled_state}. In the following, we will refer to the signal photon $(S)$ as $B$ and the idler photon $(I)$ as $X$, corresponding to the photons from the biexciton and exciton transition of a semiconductor quantum dot. The cascaded emission in a quantum dot creates a pair of photons that is energetically distinguishable due to a biexciton binding energy of typically a few millielectronvolts. These systems have been used in multiple experimental and theoretical studies on time-bin entanglement and multiple approaches exist to create time-bin entangled photon states from quantum dots \cite{jayakumar2014timebin,pril2018hyper,gines2021cavity,simon2005creating,pathak2011coherent}. However, the same equations hold true for any photon pair source.
\subsection{Quantum state tomography}
\noindent{}We define the basis states for the two-photon systems via the the single time-bin encoded states  $\Psi,\Phi\in\{E,L\}$, such that the time-bin two-photon basis reads $\ket{\Psi_B}\otimes \ket{\Phi_X} = \ket{\Psi\Phi}.$ This results in a four dimensional density matrix 
\begin{align}
\begin{split}
        \rho &= \begin{pmatrix}
        \rho_{EE,EE} & \rho_{EE,EL} & \rho_{EE,LE} & \rho_{EE,LL}\\
        \rho_{EL,EE} & \rho_{EL,EL} & \rho_{EL,LE} & \rho_{EL,LL}\\
        \rho_{LE,EE} & \rho_{LE,EL} & \rho_{LE,LE} & \rho_{LE,LL}\\
        \rho_{LL,EE} & \rho_{LL,EL} & \rho_{LL,LE} & \rho_{LL,LL}\\
    \end{pmatrix}\\
    &= 
    \frac{1}{4}\sum_{j,k=0}^{3} S_{j,k} \sigma_j\otimes\sigma_k.\label{eq:QST_DM_twophoton}
\end{split}
\end{align}
with the Stokes parameters $S_{j,k}$. As in the single-photon case, they contain the counts for measurements in different bases and can be expressed as $S_{j,k} = \mathrm{Tr}\left[(\sigma_j\otimes\sigma_k)\rho\right]$. 
Using the distinct single-photon states from Sec.~\ref{sec:1ph_tomo}, we define the states $\ket{\psi_i}=\{\ket{E},\ket{X^+},\ket{Y^+},\ket{E}\}$ and their orthogonal counterparts $\ket{\psi_{i\perp}}=\{\ket{L},\ket{X^-},\ket{Y^-},\ket{L}\}$ in this specific order, such that the Stokes parameters read \cite{altepeter2005photonic}
\begin{align}
\begin{split}
    S_{j,k} &= \left(\braket{\ket{\psi_j}\bra{\psi_{j}}}\pm\braket{\ket{\psi_{j\perp}}\bra{\psi_{j\perp}}}\right)\\
    &\otimes\left(\braket{\ket{\psi_k}\bra{\psi_{k}}}\pm\braket{\ket{\psi_{k\perp}}\bra{\psi_{k\perp}}}\right).
\end{split}\label{eq:two_qubit_stokes}
\end{align}
Here, the plus sign is used for the index $j$ or $k$ being zero and the minus sign is used otherwise. Accordingly, in the Stokes parameters, expectation values like $\braket{\ket{EE}\bra{EE}}$ or $\braket{\ket{EL}\bra{EL}}$ occur.

These expectation values correspond to coincidence measurements between the signal and idler (or biexciton and exciton) photon after they traverse a measurement setup as shown in Fig.~\ref{fig:full_measurement_setup}. The photons are split (for example by their wavelength) and sent through a set of two unbalanced measurement interferometers. A phase $\phi_X$ or $\phi_B$ can be added to the respective photon going through the long arm of the interferometer. The coincidences are then measured between the exciton and the biexciton detectors. In experiments, events at all four detectors can be recorded simultaneously. 

Following the same arguments as for the single-photon case, for the theory it is sufficient to consider coincidences between one biexciton and one exciton channel only. 
\subsection{Second-order correlation functions}
\noindent{}To account for two photons in the correlation function, we now have to rely on two-time correlation functions (or second-order correlation functions). A naive assumption would be, that because of the cascaded emission process, the $B$-photon usually arrives before the $X$-photon, resulting in $
    G^{(2)}(t,\tau) = \braket{a_{B}^{\dagger}(t)a_{X}^{\dagger}(t+\tau)a_X(t+\tau)a_B(t)}. \label{eq:twotime_bxxb}$
Here, $a_B^{\dagger}$/$a_B$ are the creation/annihilation operators of the photon emitted via the decay of the biexciton into the exciton and $a_X^{\dagger}$/$a_X$ are the creation/annihilation operators for photons generated by the exciton decay. The correlation function corresponds to the probability of a $B$-photon being detected at time $t$ and the $X$-photon being detected after delay $\tau$. However, this naive correlation function does not capture all possible events if $\tau\ge0$. Due to the delay of the interferometer, the $X$-photon might arrive at the detector before the $B$-photon does. And even if all photons take the same (short or long) paths through the interferometer, due to imperfect excitation protocols there might still be an $X$-photon sent out in the early time-bin, with a $B$-photon following in the second time-bin, corresponding to the $\ket{LE}$ state. Note that in experiments, this is usually solved by measuring coincidences of the exciton and biexciton channel with respect to a trigger pulse, for example the first excitation pulse.

Thus, it is crucial to include all possible events using the correct time ordering in the two-time correlation function as well as detecting each event in the time-spans $t_B, t_X \in [0,...,3T]$. The impact of the interferometer, inducing a delay $T$ and a phase $\phi_{r}$, is again taken into account by splitting the mode operators, resulting in
\begin{equation}
    a_r(t)\rightarrow a_r(t) + e^{i\phi_r}a_r(t-T), \quad r\in\{B,X\}. \label{eq:a_delay}
\end{equation}
Then all possible coincidences are described within the two-time correlation function
\begin{equation}
    G^{(2)}(t_B,t_X)\!=\!\left\langle\!\mathcal{T}^{-}\!\!\left[a_{B}^{\dagger}(t_B)a_{X}^{\dagger}(t_X)\right]\!\mathcal{T}^{+}\!\!\left[a_X^{\phantom{\dagger}}(t_X)a_B^{\phantom{\dagger}}(t_B)\right]\!\right\rangle.\label{eq:twotime_bxxb_timeordered}
\end{equation}
The time-ordering operators $\mathcal{T}^{\pm}$ order the smallest time to the left $(\mathcal{T}^{-})$ or to the right $(\mathcal{T}^{+})$. In total, this leads to 16 terms, each with generally two different time-orderings depending on $(t_B,t_X)$, which are given in the appendix.

\subsection{Two-time histogram}
Let us now turn to the simulations of the  measurements that are performed during the quantum state tomography. The detector counts result in Stokes parameters in Eq.~\eqref{eq:two_qubit_stokes}, that contain expectation values of the form $\braket{\ket{\psi_j\psi_k}\bra{\psi_j\psi_k}}$, where $\psi_j,\psi_k$ can be any of the six states given in Fig.~\ref{fig:onephoton_measurement_setup}(c). 
From the Stokes parameters, analogous to Sec.~\ref{sec:1ph_sim}, we calculate selected expectation values as needed for the quantum state tomography via the second order correlation function
\begin{equation}
    \braket{\ket{\Psi_B\Phi_X}\bra{\Psi_B\Phi_X}} \propto \int\limits_{\tau_1}^{\tau_1+T} \!\!\!dt_1\int\limits_{\tau_2}^{\tau_2+T} \!dt_2  \, G^{(2)}(t_1,t_2,\phi_B,\phi_X),
\end{equation}
where the limits of the integral (with $\tau_i=0,T$ or $2T$) determine the time window and how many of the 16 terms resulting from Eq.~\eqref{eq:16_terms} contribute. In experiments, these measurements automatically result in a two-time histogram as in Fig.~\ref{fig:full_measurement_setup}(b), arranged in a three-by-three grid based on their arrival times. We can group these peaks into three categories: the corner peaks, the side peaks and the center peak.

The corner peaks correspond to expectation values of the time-basis states $\braket{\ket{EE}\bra{EE}}$, $\braket{\ket{EL}\bra{EL}}$, $\braket{\ket{LE}\bra{LE}}$, $\braket{\ket{LL}\bra{LL}}$. We remember that in the two-photon basis for example $\ket{EE}$ resembles an arrival time of $t_B=0$ and $t_X=0$, i.e., both an early biexciton photon and an early exciton photon took the short path through the interferometer. In experiments this is a coincidence measurement between the $\ket{E}$-peaks that emerge for the individual measurements of biexciton and exciton channel, which was shown in Fig.~\ref{fig:onephoton_measurement_setup}(b). In the calculation, only a single term of the two-time correlation function contributes, as we integrate only over a single time window.
\begin{align}
    \braket{\ket{EE}\bra{EE}} &= \int_0^T \, dt_1\int_0^T \, dt_2 \braket{a_{B}^{\dagger}(t_1) a_{X}^{\dagger}(t_2) a_{X}(t_2) a_{B}(t_1)}
\end{align}
The other expectation values are calculated analogously and given explicitly in the appendix. 

The side peaks are states that consist of one photon being in the time-basis and the other photon being in a superposition $\ket{\Phi}$. The four peaks correspond to $\braket{\ket{E\Phi}\bra{E\Phi}}$, $\braket{\ket{\Phi L}\bra{\Phi L}}$, $\braket{\ket{L \Phi}\bra{L\Phi}}$, $\braket{\ket{\Phi E}\bra{\Phi E}}$. Let us look for example at the central left peak, which corresponds to $\braket{\ket{EX^+}\bra{EX^+}}$. It reflects the coincidence measurement between the early peak for the biexciton and the central interference peak of $\ket{X^+}$ for the exciton. As explicitly written in the appendix, for every side peak four terms of the two-time correlation function contribute to the integral. Changing just the phase $\phi_X$ allows a visibility measurement by evaluating $\braket{\ket{E\Phi}\bra{E\Phi}}$. Expanding this value results in
\begin{align}
    &\braket{\ket{E\Phi}\bra{E\Phi}} \nonumber \\
    &= \frac{1}{2}(\rho_{EE,EE} + \rho_{EL,EL} + 2\cos(\phi_X-\varphi)|\rho_{EL,EE}|).
\end{align}
This peak shows a visibility for a state like $\sim\ket{EE}+e^{i\phi}\ket{EL}$, which is however not an entangled but a classically correlated state.

The center peak shows the most complex behavior as it corresponds to a coincidence detection of the interference peaks of both biexciton and exciton channel, depending on both interferometer phases $\phi_B$ and $\phi_X$. For example, for $\phi_B=\phi_X=0$ the obtained peak is proportional to $\braket{\ket{X^+X^+}\bra{X^+X^+}}$. The measurements when turning both phase plates can be mapped to the expectation values, resulting in
\begin{align}
    \begin{split}
    &\braket{\ket{\Psi\Phi}\bra{\Psi\Phi}} \\
    &= \rho_{EE,EE} + \rho_{EL,EL} + \rho_{LE,LE} + \rho_{LL,LL}\\
    &+ 2\cos(\phi_X)|\rho_{LE,LL}| + 2\cos(\phi_X)|\rho_{EE,EL}|\\
    &+ 2\cos(\phi_B)|\rho_{EE,LE}| + 2\cos(\phi_B)|\rho_{EL,LL}|\\
    &+2\cos(\phi_B+\phi_X)|\rho_{EE,LL}| + 2\cos(\phi_B-\phi_X)|\rho_{EL,LE}| \,.
    \end{split}
\end{align}
In this peak, both phases $\phi_B$ and $\phi_X$ contribute to the measurement of the visibility. 
\subsection{Histogram projection}
In time-bin measurements, in addition to the two-time histogram, its projection along the diagonal lines, to a single time axis $t = t_B+t_X$ \cite{jayakumar2014timebin,thiel2023timebin} is often provided. The result of such a projection is depicted in Fig.~\ref{fig:full_measurement_setup}(c), showing a total of five peaks. Here, the outer peaks correspond to  $\braket{\ket{EE}\bra{EE}}$ and $\braket{\ket{ LL}\bra{LL}}$, while all other peak show variable visibilities depending on the interferometer phases. 

While this is a compact way of representing the data, now the central peak no longer serves as an indicator for the same visibility as before. Due to the projection along the diagonal, now the elements $\rho_{EL,EL}$ and $\rho_{LE,LE}$ contribute twice. Thus, this representation of the data is only useful if these peaks are relatively small, which might be the case for the time-bin entangled state given in Eq.~\eqref{eq:entangled_state}. However, for the entangled state $\ket{\psi}=\frac{1}{\sqrt{2}}(\ket{EL}+\ket{LE})$ this projection is misleading, but instead the projection along the antidiagonals would result in a similar picture.


\subsection{Time-bin entanglement}
Finally, we  discuss the usage of the measurements for identifying time-bin entangled photon pairs. In the experiment, often two-time histrograms including the corresponding visibilities of the side and the central peak are measured (i.e., the data is recorded for many settings of the phase plates). From this, the full density matrix can be reconstructed via the Stokes parameter [cf. Eq.~\eqref{eq:QST_DM_twophoton}]. 

Using the early and late operators for both exciton and biexciton and using the same procedure as in Section~\ref{sec:classification}, the density matrix can be calculated via
\begin{widetext}
    \begin{equation}
    \rho_{ij,kl} = \frac{\overline{G}^{(2)}_{ij,kl}}{\text{Tr}\left\{\overline{G}^{(2)}\right\}}, \quad \overline{G}^{(2)}_{ij,kl} = \int\limits_{0}^{T}\!dt_1\int\limits_{0}^{T}\!dt_2 \left\langle\mathcal{T}^{-}\left[a^{k\,\dagger}_{B}(t_1)a^{l\,\dagger}_{X}(t_2)\right]\mathcal{T}^{+}\left[a^j_{X}(t_2)a^i_{B}(t_1)\right]\right\rangle, \,\, i,j,k,l \in \{E,L\}.
\end{equation}
\end{widetext}
The late operators include a time-shift by $T$, i.e., $a^L_{B/X}(t) = a_{B/X}(t+T)$, while the early operators leave the time argument unchanged. See also the appendix for more information on how the different measurements are connected to the density matrix. From the density matrix a measure of entanglement, for example the concurrence \cite{cygorek2018comparison}, can be calculated. 

In certain cases, however, it is possible to extract information from a reduced set of measurements as discussed in the following:

Let us start with the case of a perfectly time-bin entangled state given in Eq.~\eqref{eq:entangled_state}.
In this case, the expectation values $\braket{\ket{EL}\bra{EL}}$ and $\braket{\ket{LE}\bra{LE}}$ as well as their corresponding coherences are zero. The only peak which then shows a visibility, i.e., a dependence on the phase of the interferometers, is the central peak of the two-time histogram, with $V=1$. However, two main factors can hinder the creation of this ideally entangled state. Firstly, the entanglement can be reduced due to a limited coherence, namely $\rho_{EE,LL}<0.5$. Secondly, $\ket{EL}$ and $\ket{LE}$ can mix into the state, resulting in finite elements $\rho_{EL,EL}$ and $\rho_{LE,LE}$, while their respective coherences remain small \cite{thiel2023timebin,jayakumar2014timebin}. In this case, the coincidence counts of the central peak are given by
\begin{equation}
    P_{\text{c}} \propto 2\cos^2\left(\frac{\phi_B+\phi_X-\varphi}{2}\right)|\rho_{EE,LL}| + \rho_{EL,EL} + \rho_{LE,LE} \, .
\end{equation}
Therefore, with the assumption that we can restrict ourselves to the aforementioned states, a reduction to the measurement of the central peak is enough to obtain a measure of the degree of entanglement. However, the main source of any deviation from the entangled state can only be identified using the full state tomography, reconstructing the density matrix of the entangled state. Including the aforementioned assumptions, also the concurrence can be approximated as \cite{seidelmann2019from}
\begin{equation}
    C \approx 2|\rho_{EE,LL}|-\rho_{EL,EL}-\rho_{LE,LE}.
\end{equation}



\section{Conclusions}
In conclusion, step-by-step we have derived the equations to calculate the density matrix of time-bin encoded photon pairs. Our derivation connects the counts that are collected in quantum state tomographic measurements to the entries of the density matrix, giving an in-depth understanding about the processes involved, revealing details of the influence of the interference terms on the measurements. This opens up the possibility to theoretically assess and optimize preparation schemes for the production of time-bin entangled photon pairs, for example by using dark states in quantum dots \cite{simon2005creating,luker2015direct,neumann2021optical, kappe2024keeping}, while simultaneously taking into account any loss or decoherence processes that can be modeled for the particular system, for example phonon influence in a semiconductor quantum dot \cite{reiter2019distinctive}. This will allow for a better evaluation of the robustness of time-bin entangled photons for future applications. 
\section{Acknowledgements}
\noindent{}We thank Stefan Frick for fruitful discussion. TKB and DER
acknowledge financial support from the German Research Foundation DFG through project 428026575 (AEQuDot). FK, YK, VR, and GW acknowledge the financial support through the Austrian Science Fund FWF projects with grant IDs 10.55776/TAI556 (DarkEneT), 10.55776/W1259 (DK-ALM Atoms, Light, and Molecules), 10.55776/FG5, 10.55776/I4380 (AEQuDot) and FFG.
\bibliography{bibfile}
\appendix
\begin{widetext}
\section{Expanded equations of the two-time correlation function}
Taking the two-time correlation function from the main text without the time-ordering operators, i.e., 
\begin{equation}
    G^{(2)}(t_1,t_2)=\braket{a_{B}^{\dagger}(t_1)a_{X}^{\dagger}(t_2)a_X^{\phantom{\dagger}}(t_2)a_B^{\phantom{\dagger}}(t_1)},
\end{equation}
and introducing the delay term
\begin{equation}
    a_r(t)\rightarrow a_r(t) + e^{i\phi_r}a_r(t-T), \quad r\in\{B,X\}, \label{eq:a_delay_app}
\end{equation}
results in a total of 16 terms. The correct time-ordering still has to be applied to each term individually, depending on $t_1,t_2$. The 16 terms are:
\begin{align}
\begin{split}
    G^{(2)}(t_1,t_2) = &e^{i \phi_{B}} e^{i \phi_{X}} \braket{a_{B}^{\dagger}(t_1) a_{X}^{\dagger}(t_2) a_{X}(t_2-T) a_{B}(t_1-T)} \\
+ &e^{i \phi_{B}} \braket{a_{B}^{\dagger}(t_1) a_{X}^{\dagger}(t_2) a_{X}(t_2) a_{B}(t_1-T)} \\
+ &e^{i \phi_{B}} \braket{a_{B}^{\dagger}(t_1) a_{X}^{\dagger}(t_2-T) a_{X}(t_2-T) a_{B}(t_1-T)} \\
+ &e^{i \phi_{B}} e^{- i \phi_{X}} \braket{a_{B}^{\dagger}(t_1) a_{X}^{\dagger}(t_2-T) a_{X}(t_2) a_{B}(t_1-T)} \\
+ &e^{i \phi_{X}} \braket{a_{B}^{\dagger}(t_1) a_{X}^{\dagger}(t_2) a_{X}(t_2-T) a_{B}(t_1)} \\
+ &e^{i \phi_{X}} \braket{a_{B}^{\dagger}(t_1-T) a_{X}^{\dagger}(t_2) a_{X}(t_2-T) a_{B}(t_1-T)} \\
+ &\braket{a_{B}^{\dagger}(t_1) a_{X}^{\dagger}(t_2) a_{X}(t_2) a_{B}(t_1)} \\
+ &\braket{a_{B}^{\dagger}(t_1) a_{X}^{\dagger}(t_2-T) a_{X}(t_2-T) a_{B}(t_1)} \\
+ &\braket{a_{B}^{\dagger}(t_1-T) a_{X}^{\dagger}(t_2) a_{X}(t_2) a_{B}(t_1-T)} \\
+ &\braket{a_{B}^{\dagger}(t_1-T) a_{X}^{\dagger}(t_2-T) a_{X}(t_2-T) a_{B}(t_1-T)} \\
+ &e^{- i \phi_{X}} \braket{a_{B}^{\dagger}(t_1) a_{X}^{\dagger}(t_2-T) a_{X}(t_2) a_{B}(t_1)} \\
+ &e^{- i \phi_{X}} \braket{a_{B}^{\dagger}(t_1-T) a_{X}^{\dagger}(t_2-T) a_{X}(t_2) a_{B}(t_1-T)} \\
+ &e^{- i \phi_{B}} e^{i \phi_{X}} \braket{a_{B}^{\dagger}(t_1-T) a_{X}^{\dagger}(t_2) a_{X}(t_2-T) a_{B}(t_1)} \\
+ &e^{- i \phi_{B}} \braket{a_{B}^{\dagger}(t_1-T) a_{X}^{\dagger}(t_2) a_{X}(t_2) a_{B}(t_1)} \\
+ &e^{- i \phi_{B}} \braket{a_{B}^{\dagger}(t_1-T) a_{X}^{\dagger}(t_2-T) a_{X}(t_2-T) a_{B}(t_1)} \\
+ &e^{- i \phi_{B}}e^{- i \phi_{X}}  \braket{a_{B}^{\dagger}(t_1-T) a_{X}^{\dagger}(t_2-T) a_{X}(t_2) a_{B}(t_1)}.
\end{split}
\label{eq:16_terms}
\end{align}
These 16 terms can then be connected to the measurements in the 9 peaks that were shown in Fig.~\ref{fig:full_measurement_setup}(b) in the main text. We assume that the expectation values of the operators in Eq.~\eqref{eq:16_terms} are only unequal to zero if the time argument is within $0,...,2T$. This simplification is exact under the assumption that at least in the time-bin before and after the excitation scheme that prepares the time-bin encoded state, no other pulse excites the system, and that the system fully relaxes during the period of one time-bin (i.e., it does not leak into a third time-bin). This lets us select only the relevant terms for each peak in the measurement and was implicitly also done for the calculations of only one photon.\\
Following the same procedure as in Section~\ref{sec:classification} in the main text and using the Stokes parameters for the two-qubit case, we arrive at the density matrix
\begin{equation}
    \rho = \frac{\overline{G}^{(2)}_{ij,kl}}{\text{Tr}\left\{\overline{G}^{(2)}\right\}}, \quad \overline{G}^{(2)}_{ij,kl} = \int\limits_{0}^{T}\!dt_1\int\limits_{0}^{T}\!dt_2 \left\langle\mathcal{T}^{-}\left[a^{k\,\dagger}_{B}(t_1)a^{l\,\dagger}_{X}(t_2)\right]\mathcal{T}^{+}\left[a^j_{X}(t_2)a^i_{B}(t_1)\right]\right\rangle, \,\, i,j,k,l \in \{E,L\}.
\end{equation}
In the following, we give further information on the distinct peaks of the two-dimensional histogram and how they can be understood and calculated. 
For the four corners of the diagram, corresponding to the states of the combined time-basis, it follows that
\begin{subequations}
\begin{align}
    \braket{\ket{EE}\bra{EE}} &= \int_0^T \!dt_1\int_0^T \!dt_2 \braket{a_{B}^{\dagger}(t_1) a_{X}^{\dagger}(t_2) a_{X}(t_2) a_{B}(t_1)}\\
    \braket{\ket{LL}\bra{LL}} &= \int_{2T}^{3T} \!dt_1\int_{2T}^{3T} \!dt_2 \braket{a_{B}^{\dagger}(t_1-T) a_{X}^{\dagger}(t_2-T) a_{X}(t_2-T) a_{B}(t_1-T)}\\
    \braket{\ket{EL}\bra{EL}} &= \int_0^T \!dt_1\int_{2T}^{3T} \!dt_2 \braket{a_{B}^{\dagger}(t_1) a_{X}^{\dagger}(t_2-T) a_{X}(t_2-T) a_{B}(t_1)}\\
    \braket{\ket{LE}\bra{LE}} &= \int_{2T}^{3T} \!dt_1\int_0^T \!dt_2 \braket{a_{B}^{\dagger}(t_1-T) a_{X}^{\dagger}(t_2) a_{X}(t_2) a_{B}(t_1-T)}
\end{align}
\end{subequations}
Remember that the correct time-ordering has to be satisfied in the correlation functions of these equations.\\
For the combinations of time- and energy basis the phases of one of the interferometers enters. For example, the correlation of the early peak of the biexciton and the middle peak of the exciton, for $\ket{\Phi}\in\{\ket{X^{\pm}},\ket{Y^{\pm}}\}$ depending on phase $\phi_X$, can be calculated using
\begin{align}
\begin{split}
    \braket{\ket{E\Phi}\bra{E\Phi}} \propto \frac{1}{2} \int_0^T \!dt_1\int_T^{2T} \!dt_2 &\phantom{+}(\braket{a_{B}^{\dagger}(t_1) a_{X}^{\dagger}(t_2) a_{X}(t_2) a_{B}(t_1)}\\
    &+ \braket{a_{B}^{\dagger}(t_1) a_{X}^{\dagger}(t_2-T) a_{X}(t_2-T) a_{B}(t_1)}\\
    &+ e^{i \phi_{X}} \braket{a_{B}^{\dagger}(t_1) a_{X}^{\dagger}(t_2) a_{X}(t_2-T) a_{B}(t_1)}\\
    &+ e^{- i \phi_{X}} \braket{a_{B}^{\dagger}(t_1) a_{X}^{\dagger}(t_2-T) a_{X}(t_2) a_{B}(t_1)})
\end{split}
\end{align}
We can expand $\braket{\ket{E\Phi}\bra{E\Phi}}$ into the time-basis to see what density matrix elements play a role in this interference peak.
\begin{align}
    \braket{\ket{E\Phi}\bra{E\Phi}} &= \frac{1}{2}(\braket{\ket{EE}\bra{EE}} + \braket{\ket{EL}\bra{EL}} + e^{i\phi_X}\braket{\ket{EL}\bra{EE}} + e^{-i\phi_X}\braket{\ket{EE}\bra{EL}}) \notag \\
    &= \frac{1}{2}(\rho_{EE,EE} + \rho_{EL,EL} + 2\cos(\phi_X-\varphi)|\rho_{EL,EE}|)
\end{align}
It is obvious that this peak also shows a visibility if the photon would be in the state $\propto\ket{EE}+e^{i\phi}\ket{EL}$, which is however, clearly no entangled state. In total, this peak can be understood as the interference between the $\ket{EE}$ and $\ket{EL}$ states.\\
The most complex behavior is present in the central peak of the two-dimensional histogram. It corresponds to a measurement in the energy basis for both the biexciton and the exciton photon, thus it is the only peak where both phases $\phi_B,\phi_X$ play a role. In fact, all of the 16 terms of Eq.~\eqref{eq:16_terms} contribute to this peak, so it follows that
\begin{equation}
    \braket{\ket{\Psi_B\Phi_X}\bra{\Psi_B\Phi_X}} \propto \int_T^{2T} \!dt_1\int_T^{2T} \!dt_2  \, G^{(2)}(t_1,t_2,\phi_B,\phi_X),
\end{equation}
where $\ket{\Phi},\ket{\Psi}\in\{\ket{X^{\pm}},\ket{Y^{\pm}}\}$ depending on the settings of the phase plates $\phi_B,\phi_X$. Expanding to the time-basis and substituting the density matrix elements (neglecting a constant phase $\phi$) leads to
\begin{align}
    \braket{\ket{\Psi\Phi}\bra{\Psi\Phi}} &= \rho_{EE,EE} + \rho_{EL,EL} + \rho_{LE,LE} + \rho_{LL,LL} \notag \\
    &+ 2\cos(\phi_X)|\rho_{LE,LL}| + 2\cos(\phi_X)|\rho_{EE,EL}| \notag \\
    &+ 2\cos(\phi_B)|\rho_{EE,LE}| + 2\cos(\phi_B)|\rho_{EL,LL}| \notag\\
    &+2\cos(\phi_B+\phi_X)|\rho_{EE,LL}| + 2\cos(\phi_B-\phi_X)|\rho_{EL,LE}|
\end{align}
In principle, this peak can be understood as an interference between all possible time-bin states. Most importantly, it also includes the coherences $\rho_{EE,LL}$ and $\rho_{EL,LE}$. If the time-bin entanglement is only limited by, for example, a limited coherence $\rho_{EE,LL}$ while all other coherences are zero, the visibility reduces to $V=2\rho_{EE,LL}$ which is unity in the ideal case. 
\section{Correlation functions for the calculation of the density matrix}
Let us recall the general formula for the two-time correlation functions that is given in the main text in Eq.~\eqref{eq:QST_DM_twophoton}. For calculating the time-bin entanglement for photons coming directly from a semiconductor quantum dot, we substitute the photon creation (annihilation) operators $a^{\dagger}_{B/X} (a_{B/X})$ with the quantum dots' transition operators $\sigma_{B/X}^{\dagger} (\sigma_{B/X})$:
\begin{equation}
    \overline{G}^{(2)}_{ij,kl} = \int\limits_{0}^{T}\!dt_1\int\limits_{0}^{T}\!dt_2 \braket{\mathcal{T}^{-}\left[\sigma^{k\,\dagger}_{B}(t_1)\sigma^{l\,\dagger}_{X}(t_2)\right]\mathcal{T}^{+}\left[\sigma^j_{X}(t_2)\sigma^i_{B}(t_1)\right]},
\end{equation}
From this, we will derive the formulas for the ten different entries of the two-photon density matrix. We start with $\overline{G}^{(2)}_{EE,EE}$:
\begin{equation}
    \overline{G}^{(2)}_{EE,EE} = \int\limits_{0}^{T}\!dt_1\int\limits_{0}^{T}\!dt_2 \braket{\mathcal{T}^{-}\left[\sigma^{\dagger}_{B}(t_1)\sigma^{\dagger}_{X}(t_2)\right]\mathcal{T}^{+}\left[\sigma_{X}(t_2)\sigma_{B}(t_1)\right]},
\end{equation}
We will split this equation in two parts that have a different time-ordering: (a) $t_1 \le t_2$ and (b) $t_2 \le t_1$. 
\begin{equation}
    \overline{G}^{(2)}_{EE,EE} = \underbrace{\int\limits_{0}^{T}\!dt_1\int\limits_{t1}^{T}\!dt_2 \braket{\sigma^\dagger_{B}(t_1)\sigma^\dagger_{X}(t_2)\sigma_{X}(t_2)\sigma_{B}(t_1)}}_{\text{(a)}}
     + \underbrace{\int\limits_{0}^{T}\!dt_2\int\limits_{t2}^{T}\!dt_1 \braket{\sigma^\dagger_{X}(t_2)\sigma^\dagger_{B}(t_1)\sigma_{B}(t_1)\sigma_{X}(t_2)}}_{\text{(b)}}
\end{equation}
Now, in part (b) renaming $t_1\rightarrow t_2$ and simultaneously $t_2 \rightarrow t_1$ allows to put it into the same format as (a), only with interchanged operators. Note that the special case $t_1=t_2$ is included in both integrals (a) and (b), but in (b) this always results in a zero contribution, as with identical time arguments $\sigma_B\sigma_X=0$. This renaming of the time arguments will also be done in the following to bring the equations into a similar form. $\overline{G}^{(2)}_{LL,LL}$ has the same form but is shifted in time by $T$:
\begin{align}
    \overline{G}^{(2)}_{LL,LL} &= \int\limits_{0}^{T}\!dt_1\int\limits_{t1}^{T}\!dt_2 \braket{\sigma^\dagger_{B}(t_1+T)\sigma^\dagger_{X}(t_2+T)\sigma_{X}(t_2+T)\sigma_{B}(t_1+T)}  \notag\\
     &+ \int\limits_{0}^{T}\!dt_1\int\limits_{t1}^{T}\!dt_2 \braket{\sigma^\dagger_{X}(t_1+T)\sigma^\dagger_{B}(t_2+T)\sigma_{B}(t_2+T)\sigma_{X}(t_1+T)}
\end{align}
For $\overline{G}^{(2)}_{EL,EL}$ and $\overline{G}^{(2)}_{LE,LE}$ the arguments are always separated in time by $T$, meaning the distinction of the two parts is not necessary:
\begin{align}
    \overline{G}^{(2)}_{EL,EL} &= \int\limits_{0}^{T}\!dt_1\int\limits_{0}^{T}\!dt_2 \braket{\sigma^\dagger_{B}(t_1)\sigma^\dagger_{X}(t_2+T)\sigma_{X}(t_2+T)\sigma_{B}(t_1)}, \\
    \overline{G}^{(2)}_{LE,LE} &= \int\limits_{0}^{T}\!dt_1\int\limits_{0}^{T}\!dt_2 \braket{\sigma^\dagger_{X}(t_1)\sigma^\dagger_{B}(t_2+T)\sigma_{B}(t_2+T)\sigma_{X}(t_1)}.
\end{align}
The only difference between these two is the ordering of the operators, i.e., interchanging $X\leftrightarrow B$. Similar symmetries can also be found for $EL,EE$ and $LE,EE$:
\begin{align}
    \overline{G}^{(2)}_{EL,EE} &= \int\limits_{0}^{T}\!dt_1\int\limits_{t1}^{T}\!dt_2 \braket{\sigma^\dagger_{B}(t_1)\sigma^\dagger_{X}(t_2)\sigma_{X}(t_2+T)\sigma_{B}(t_1)} \notag \\
     &+ \int\limits_{0}^{T}\!dt_1\int\limits_{t1}^{T}\!dt_2 \braket{\sigma^\dagger_{X}(t_1)\sigma^\dagger_{B}(t_2)\sigma_{X}(t_1+T)\sigma_{B}(t_2)},\\
    \overline{G}^{(2)}_{LE,EE} &= \int\limits_{0}^{T}\!dt_1\int\limits_{t1}^{T}\!dt_2 \braket{\sigma^\dagger_{B}(t_1)\sigma^\dagger_{X}(t_2)\sigma_{B}(t_1+T)\sigma_{X}(t_2)}  \notag\\
     &+ \int\limits_{0}^{T}\!dt_1\int\limits_{t1}^{T}\!dt_2 \braket{\sigma^\dagger_{X}(t_1)\sigma^\dagger_{B}(t_2)\sigma_{B}(t_2+T)\sigma_{X}(t_1)}.
\end{align}
Here, interchanging $X\leftrightarrow B$ and (a) $\leftrightarrow$ (b) leads to the same form of the equation.
The equations for $LL,EL$ and $LL,LE$ read
\begin{align}
    \overline{G}^{(2)}_{LL,EL} &= \int\limits_{0}^{T}\!dt_1\int\limits_{t1}^{T}\!dt_2 \braket{\sigma^\dagger_{B}(t_1)\sigma^\dagger_{X}(t_2+T)\sigma_{X}(t_2+T)\sigma_{B}(t_1+T)} \notag \\
     &+ \int\limits_{0}^{T}\!dt_1\int\limits_{t1}^{T}\!dt_2 \braket{\sigma^\dagger_{B}(t_2)\sigma^\dagger_{X}(t_1+T)\sigma_{B}(t_2+T)\sigma_{X}(t_1+T)},\\
    \overline{G}^{(2)}_{LL,LE} &= \int\limits_{0}^{T}\!dt_1\int\limits_{t1}^{T}\!dt_2 \braket{\sigma^\dagger_{X}(t_2)\sigma^\dagger_{B}(t_1+T)\sigma_{X}(t_2+T)\sigma_{B}(t_1+T)}\notag \\
     &+ \int\limits_{0}^{T}\!dt_1\int\limits_{t1}^{T}\!dt_2 \braket{\sigma^\dagger_{X}(t_1)\sigma^\dagger_{B}(t_2+T)\sigma_{B}(t_2+T)\sigma_{X}(t_1+T)}.
\end{align}
For these, interchanging $X\leftrightarrow B$ and (a) $\leftrightarrow$ (b) leads to the same form of the equation as well.\\
Up to this point, the equations consisted of two-time correlations for the diagonal of the density matrix and three-time correlations for the off-diagonal parts. The two remaining parts on the anti-diagonal, $LE,EL$ and $LL,EE$, result in four-time correlation functions:
\begin{align}
    \overline{G}^{(2)}_{LE,EL} &= \int\limits_{0}^{T}\!dt_1\int\limits_{t1}^{T}\!dt_2 \braket{\sigma^\dagger_{B}(t_1)\sigma^\dagger_{X}(t_2+T)\sigma_{B}(t_1+T)\sigma_{B}(t_2)}\notag \\
     &+ \int\limits_{0}^{T}\!dt_1\int\limits_{t1}^{T}\!dt_2 \braket{\sigma^\dagger_{B}(t_2)\sigma^\dagger_{X}(t_1+T)\sigma_{B}(t_2+T)\sigma_{X}(t_1)},\\
    \overline{G}^{(2)}_{LL,EE} &= \int\limits_{0}^{T}\!dt_1\int\limits_{t1}^{T}\!dt_2 \braket{\sigma^\dagger_{B}(t_1)\sigma^\dagger_{X}(t_2)\sigma_{X}(t_2+T)\sigma_{B}(t_1+T)}\notag \\
     &+ \int\limits_{0}^{T}\!dt_1\int\limits_{t1}^{T}\!dt_2 \braket{\sigma^\dagger_{X}(t_1)\sigma^\dagger_{B}(t_2)\sigma_{B}(t_2+T)\sigma_{X}(t_1+T)}.
\end{align}
With these equations, the two-photon density matrix for photons from a quantum dot can be calculated.
\end{widetext}
\end{document}